\documentstyle[12pt,psfig]{article}
\begin{document}
\vskip 0.5 true cm
{\bf  \Large Fermi Statistics of Weakly Excited Granular Materials in
a Vibrating Bed: Molecular Dynamics Simulations}
\vskip 0.5 true cm
\centerline {Paul V. Quinn \thanks{E-mail address: pvq2@lehigh.edu}
and Daniel C. Hong \thanks{
E-mail address: dh09@lehigh.edu}}
\vskip 0.3 true cm
\centerline {\bf Department of Physics, Lewis Laboratory, Lehigh University,
Bethlehem, Pennsylvania 18015}

\date{\today}

\begin{abstract}
Molecular Dynamics simulations were carried out to test the
thermodynamic theory of weakly excited, two-dimensional,
granular systems [Hayakawa and Hong, Phys. Rev. Lett. {\bf 78}, 2764(1997)],
where granular materials are viewed as a collection of spinless Fermions.
We first determine
the global temperature $T$ by fitting the steady state 
density profile to the Fermi distribution function, and then
measure the center of mass,
$<z(T)>$, and its fluctuations, $ <(\Delta z(T))^2>$ 
as a function of T.  We find a fairly good agreement
between theory and simulations, in particular, in the estimation of
the temperature and the scaling behavior of $<z(T)>$ and $<(\Delta z(T))^2>$.
\end{abstract}
\vskip 0.2 true cm
\noindent PACS numbers: 81.05 Rm, 05., 07.05Tp, 82.20.wt

\noindent Key Words: Granular materials, excluded volume effect, Fermi
statistics.
\vskip 2.0 true cm
\noindent {\bf I. Introduction}
\vskip 0.2 true cm
In a recent paper, Hayakawa and Hong(HH) [1] 
explored a simple consequence of the 
excluded volume effect for a dense granular system.  They proposed a 
thermodynamic theory for weakly
excited, dissipative, nonequilibrium, granular
systems from the view point of the elementary excitation of non-interacting
spinless Fermions.   Based on a simple thermodynamic argument and
combined with the observation that 
the granular state in 
a vibrating bed may be viewed as
an excited state,
HH demonstrated that the steady state density profile must be given by
the Fermi distribution function.  This  
enables one to associate and define 
a macroscopic, global
parameter, T, which is
suitable to characterize the state of the vibrating
system.  The parameter T is the thermodynamic temperature, and
its relation to external control parameters has been derived in ref.[1].
The theory of Hayakawa and Hong has been successful in
explaining the observed density profile of the experimental data of
Clement and Rajchenbach [2].  
The purpose of this
paper is to further
test the predictions of HH using Molecular Dynamics simulations, in particular,
 with regards to the configurational statistics
of granular materials in a vibrating bed.  
We first briefly explain the background of the thermodynamic theory in 
section II along with a summary of the theory of ref.[1] in section III.
We then present
the simulation results in section IV.  In section V,
is a discussion of some of the important issues related to this work.
\vskip 1.0 true cm
\noindent {\bf II. Background of Fermi Statistics
and Thermodynamic Theory of Granular Materials}
\vskip 0.2 true cm
The system being studied here is a dense, dissipative, nonequilibrium, 
granular system, where the mean free path of the grains is of the 
order of a few particle
diameters.  Hence, each particle may be considered to be effectively confined
in a cage as in the free volume theory of a dense liquid [4].  In
such a case, an observation has been made in ref.[1] that the basic granular
state is not a gas, but a solid or crystal, and thus the effective
thermodynamic theory based on the free energy argument may be more
appropriate than the kinetic theory in studying this state.  In such
a case, the {\it configurational} statistics of the steady state may be
determined by the variational method as the most
probable or minimum free energy state.  

To be more specific,
consider the excitation of disordered granular materials confined
in a box with vibrations
of the bottom plate.  The vibrations
will inject energy into the system which cause the ground
state to become unstable, and a newly excited state will emerge with
an expanded volume.  
The time averaged configurational states of this new excited state
have undergone structural distortions.  However, the degree of
distortions from the ground state may be small for a weakly excited
state, possibly justifying the use of an
effective thermodynamic theory based on the variational principle.
Such a thermodynamic approach may be further
justified by the following two recent experiments:
\vskip 0.2 true cm

1. Weakly or moderately excited regime:
Clement and Rajchenbach(CR) [2]
have performed an experiment with the vibrational strength,
$\Gamma$, of the order one
for a two dimensional vibrating bed,
using inclined side walls to suppress  convections.  Here, $\Gamma=A\omega^2/g$
 with A and $\omega$, the amplitude and frequency of the vibrating plane,
and g, the gravitational constant.
CR have 
found that the ensemble-averaged density profile 
as a function of
height from the bottom layer obeys a universal function that is {\it
independent}
of the phase of oscillations of the vibrating plate. Namely, it is independent
of the kinetics imposed on the system.  One conceptually important point
here is that the reference point of the density profile is not the bottom
plate, but the bottom layer, which of course is fluidized.
\vskip 0.4 true cm
2. Highly excited regime: Warr and Hansen(WH) [5]
have performed an experiment on 
highly agitated, vertically vibrating beds of $\Gamma \approx 30-50$
using
steel balls with a small coefficient of restitution.  
They have found that the collective behaviors of this
vibrated granular medium
{\it in a  stationary nonequilibrium state} 
exhibits strong
similarities to those of an atomistic fluid
in  {\it thermal equilibrium} at the corresponding particle packing 
fraction, in particular, in the two-point correlation function [6]. 
\vskip 0.3 true cm

The results of both experiments indicate that for both
moderate or highly excited systems,
one-to-one correspondence seems to exist between {\it configurational}
statistics of the
nonequilibrium stationary state and the equilibrium thermal state.
In fact, this is not so surprising considering that upon
vibration, the granular materials expand and increase the volume of the 
system.  In turn, this increase corresponds to a rise in the potential 
energy after the configurational average is appropriately taken.  
Then the
problem reduces to the packing problem, and the temperature-like variable,
T, can be associated to the vibrating bed.
The existence of distinctive configurational statistics
in the density profile of CR (and also in WH in a special case) 
appears to be fairly convincing evidence that kinetic aspects 
of the excited granular materials
may be  separated out from the statistical configurations.
These observations are the basis of
the thermodynamic theory proposed in
[1].  Note that the Fermi statistics is essentially the macroscopic
manifestation of the classical
excluded volume effect and the anisotropy
which causes the ordering of potential energy by gravity.
The top surface of the granules plays the role of
a Fermi surface, and the thin boundary layers that appear
near the top layer upon excitation
play the role of excited electrons of the Fermi gas in metals.
\vskip 0.5 true cm
\noindent {\bf III. Thermodynamic Theory of Weakly Excited Granular
Systems}
\vskip 0.3 true cm
{\bf 1. Fermi temperature}:
In ref.[1], the vibrating system was viewed from two different points of view:
one may view it as a mechanical system, in which case the expansion is due
to excitation induced by mechanical vibrations of vibration
strength $\Gamma=A\omega^2/g$. 
In this case,
the expansion is purely due to kinetics.
  In an attempt to develop
a thermodynamic theory, such a system was
 also viewed in ref.[1] as a thermal system in contact with a heat 
reservoir, and a global temperature T was associated with it.  In this case,
the expansion is purely a thermal expansion.
By equating the thermal expansion, $\Delta h$, defined as
the increase in the center of mass,
and the kinetic expansion, $g/\omega^2 H_o(\Gamma)$, where $H_o(\Gamma)$ is
the jump height of a single ball [1,7]
at the vibrating plate(see eq.(1) in ref.[1]),
a closure 
in the thermodynamic theory of powders was obtained in ref.[1].
   Since the density decrease
above the Fermi surface is not sharp, but smooth, one may replace
$gH_o(\Gamma)/\omega^2$ with $\bar h_o(\Gamma)/\alpha$, where 
$\bar h_o(\Gamma)$
is the maximum jump height of a single ball at the Fermi surface(or 
vibrating plate) determined
by MD simulations.  
The factor $\alpha$ was introduced to incorporate (i) the smooth
decrease in the density profile near the Fermi surface, and (ii) 
the suppression in the jump height due to dissipation.  
By equating the kinetic expansion
and the thermal expansion,
one obtains the follwing explicit relationships between the temperature T 
and the control parameters:
$$ \frac{T}{mg}\qquad\qquad 
= \frac{1}{\pi}\sqrt{
\frac{6D(gH_o(\Gamma)/\omega^2)}{\alpha})}\eqno (1a)$$
$$\qquad\qquad = \frac{1}{\pi}\sqrt{\frac{6D\bar h_o(\Gamma)}{\alpha}}.
\eqno (1b)$$
Note that when a single particle is 
on a vibrating plate, the energy from the bottom wall is transferred to the 
particle through direct contact.  In the case of many particles, the supplied
energy at the vibrating plate
must first travel through other particles locked in their respective lattice 
states before reaching
those particles in the fluidized layer.  For (1a), $gH_o(\Gamma)/\omega^2$,
the maximum of the single ball jump height, is a lower bound 
because the relative velocity between the ball and the plate is assumed to be 
zero, making this temperature a lower limit for the system.  In reality the 
relative velocity can be much higher and 
$\alpha$ could be smaller than 1.  
For (1b), the expansion of the fluidized layer near the Fermi surface
is certainly less than the 
simulated single ball jump height because of the dissipation of energies 
through collisions. In this case, $\alpha$ is like a dissipation constant.  
The best fit $\alpha$'s used in this work were $\alpha=1$ in (1a) and
$\alpha$ = 64/5(fractional fit) in (1b).
The second value of $\alpha$ fit all the data, 
regardless of whether the sine wave or the triangular wave was used to vibrate 
the bottom wall.
\vskip 0.5 true cm
{\bf 2. The center of mass and its fluctuations}:
Since the density profile is given by the Fermi function, it is straightfoward
to compute the center of mass, $z(T)$, and its fluctuations,
$<(\Delta z)^2>$.  The following formulas
were derived in ref.[1]:

$$\Delta z(T) = z(T)-z(0) = \frac{D\mu_o\pi^2}{6}(\frac{T}{mgD\mu_o})^2 \eqno
(2a)$$

$$ <(\Delta z)^2> = <(z(T)-<z>)^2> = <(\Delta h)^2>/\mu_o^2
=\frac{\pi^2}{3}(\frac{T}{mgD})^3\frac{D^2}{\mu_o^2}.\eqno (2b)$$
Note that the total expansion, $\Delta h(T)\equiv \mu_o\Delta z$ and its
fluctuations $<(\Delta h)^2>/D^2 = <\mu_o(\Delta z)^2/D^2>$ are
only a function of the dimensionless Fermi temperature $T_f=T/mgD$ as 
expected.  Further,
note that (2b) is an indirect confirmation that the specific heat is linear in
T as it is for the non-interacting Fermi gas.  
\vskip 1.0 true cm

\noindent {\bf IV. Test of Fermi Statistics}
\vskip 0.3 true cm
We have carried out Molecular Dynamics simulations 
to test the thermodynamic theory of ref.[1] by specifically measuring the
density profile, its center of mass and the fluctuations of the center of mass.
The MD code used in this paper was provided by Jysoo Lee.  It is identical to 
those used by HLRZ group in Germany [8] and thus we refer to ref.[8]
for details about the code.  To start a simulation, the 
program first places the particles into a two-dimensional configuration 
specified in the code by the user.  Then the particles are allowed 
to relax under gravity and the particle-particle and particle-wall 
interactions before any vibration is ``turned on''.  The vibration of the 
bottom plate should not begin until the particles are motionless.  This 
relaxation time is found through trial and error while tracking the motion of
the particles.  The configurational
average was taken every 100 time steps, and  
$t_N \approx 2 \bullet 10^7$, which is approximately 2000 vibrating cycles.
Then, the center of mass and the fluctuations are plotted
as a function of time.  We now present the data.

\vskip 0.2 true cm
\noindent {\bf 1. The Fermi Temperature T} 

Simulations were carried out in two dimensional boxes with vertical walls
for N particles each with a diameter
$D=0.2cm$ and a mass $m=4\pi(D/2)^3/3=4.188\bullet 10^{-3}gm$ 
with degeneracy $\Omega$ 
using (N,$\Omega$) = (100,4), (200,4), and (200,8) with
a sine wave vibration and (100,4) and (200,4) with a triangular wave.
Here, degeneracy is simply the number of columns.
The dimensionless Fermi energy $\mu_o=\mu/D=N/\Omega$,
which is the initial layer thinckness.
Intially, the particles were arranged on
a square lattice, and the frequency  $\omega$ was set to a
fixed value, 
$\omega=2\pi f= 40\pi$, which is  
the experimental value (f=20Hz) used by Clement and Rajchenbach [3].  The
strength of the vibration, $\Gamma=A\omega^2/g$, was 
varied by changing the amplitude, A, of
the vibration.  
Initially, $\Gamma$ was varied from 1.2 to 20 in the simulations with 
$g=981cm/s^2$.  The density profile
closely follows the Fermi profile for low $\Gamma 's$, but it crosses over 
to the Botzmann distribution for higher $\Gamma 's$ (Fig.1).  Note that
the deviation from the Fermi profile for large $\Gamma$ is quite noticeable.  
The deviation of the $\Gamma$ = 20 data from the Boltzmann function near the
bottom plate is due to 
the model used to approximate particle-particle reactions in the simulation 
program.  The springs used in the soft sphere model of these particles  
cause the odd motion observed in the bottom layers of the columns of particles.
  These bottom particles have more of a spring force than the top layers 
because the extra mass on top causes the bottom particles to deform more.  
This deformation leads to the separation of these bottom particles from the 
bulk, giving the density profile its odd shape from $0 < z \le 4$ for 
$\Gamma$ = 20.  
Hence, our analysis is
focused strictly for $\Gamma \leq 4$.  $A/D$ changes from 0.373 for 
$\Gamma=1.2$ to 1.24 for $\Gamma=4$.  For all the data, 
$4.90 \le R=\mu D/\Gamma A < 111.20$, and thus, the Fermi statistics is 
satisfied (See ref.[1] for this criteria).

Representative samples of the density profiles
for the sine wave  with 
(N,$\Omega$)=(100,4)
for $\Gamma =1.2,1.5,2.0,2.5,3.0,3.5$ and $4.0$ are plotted in 
Fig.2.  Similar profiles were obtained for other sets of
parameters.  The data are then fitted by the Fermi function, and the  
Fermi temperature was determined. (Note that the temperature, $T_f$, 
discussed in the following 
text is actually $T/mg$. This is because the fitting function 
$\rho(z)= \rho_c/(1+exp[(z-\mu)/T_f])$ (See eq.(2))
was used with $\mu =\mu_o D$ 
and $z = (i-1/2)D $ (height with units of centimeters.)

If $\bar h_o(\Gamma)$ with $\alpha=64/5$ is used in our temperature
formula, the predicted values match extremely well with the measured ones. A
reasonable agreement between the theory and the simulation is reached
when one uses $\bar h_o(\Gamma)$ for both the sine and triangular wave data
as shown in Fig.3a and 3b.
This agreement is fairly constant for all five simulated systems.  
Now we turn our attention to
the study of the temperature scaling.

\vskip 0.2 true cm
\noindent {\bf 2. Study of temperature scaling}

We now test
the scaling relation predicted by eq.(1a) between
the temperature and control parameters such as
frequency ($\omega$), gravitational acceleration (g), 
and particle diameter (D).
\vskip 0.2 true cm
(a) {\bf Frequency Dependence}:  The frequency of the
sine wave was varied for fixed values of
$\Gamma = 2.0$, N=200, and $\Omega=4$,  over 
the range $5\leq\omega\leq35$.  The temperature was found to scale with
frequency as
$$ T \approx \omega^{m_1}$$
where $m\approx 1.16$, which is fairly close to the
predicted value $m_1=1$.  (Fig.4a)
\vskip 0.2 true cm
(b) {\bf Gravity Dependence}:  Temperature as a function of the 
gravitational acceleration is displayed in Fig.4b 
with fixed values of $\Gamma=2.0$, N=200, $\Omega=4$, and 
$f=\omega/2\pi=20 Hz$ for the range of $150\leq g \leq 1200$.  
The temperature scales as
$$ T \approx g^{-m_2}$$
where $m_2 \approx 0.48$, which is again, close to the predicted value 
$m_2=0.5$.  We notice
a change in the temperature dependence for $g_c\ge g$ with
$g_c\approx 1000 cm/s^2$, the reason for which is not so clear at this
point.
\vskip 0.2 true cm
(c) {\bf Diameter Dependence}:  The temperature as a function of the 
diameter, D, at the fixed frequency $f=\omega/2\pi=20 Hz$ and 
$g=981cm/sec^2$ is displayed in Fig.4c with fixed values of $\Gamma = 2.0$, 
N=200, and $\Omega=4$ for the range $0.025 \leq D \leq 1.60$.  Here, 
as the diameter was changed, the density was changed accordingly using 
$\rho=(3/4\pi D^3)m$ to ensure that the mass remained at a constant value of 
$4.188 x 10^{-3} gm$.  The temperature scales as
$$ T \approx D^{-m_3}$$
where $m_3 \approx 0.53$, which is again, close to 
the predicted value for $m_3=0.5$.  

To summarize, we find that the scaling relations between
the temperature and $\omega$, g, and D appear to be consistent with
the theoretical predictions within the error bars.
Our next task is to examine
the scaling relations presented for the center of mass and its fluctuations.
\vskip 0.5 true cm
{\bf 3. Center of Mass}

The center of mass is plotted in Fig. 5 
as a function of $T^2$ in a graph of seven different 
$\Gamma$'s for N=200, $\Omega=4$ and $\Omega=8$, and N=100, $\Omega=4$ under a 
sine wave vibration and N=100, $\Omega=4$ and N=200, $\Omega=4$ under the 
triangle wave vibration.  The solid line, a guide for the 
eye, seems to confirm the scaling predictions of the Fermi statistics as given
by the formula (2a), namely
$$\Delta z \propto T^2.$$
However, there exists a large discrepancy in the amplitude, C.
For N=200 and $\Omega=4$, the theory predicts
that one should get from (2a)
$C=\pi^2/6\mu_o =\pi^2/6\mu \approx .16$ with $\mu=D\mu_o$, the actual
height of the Fermi surface.  The
simulation results yield, $C\approx 3.0$ for the sine wave and $C\approx 5.6$ 
for the triangular wave, a difference of a factor 20 and greater 
when compared to with the theoretical prediction.  
This discrepancy has been traced in some detail using 
Mathematica, and it appears that the center of mass defined in (2a) is
{\it extremely} sensitive to a slight change in the Fermi energy, $\mu$.
For example, if the average $\mu=10.01$ is used for the N=200, $\Omega=4$ 
sine wave data and the integral is computed with Mathematica, the plot of 
the center of mass values as a function of $T^2$ yield a slope that {\it is} 
indeed 0.16, just as predicted by the
Fermi analysis.  However, if the experimentally determined $\mu$  
ranging from 9.91 to 10.13 is used  and the center of mass is computed using 
Mathematica, the slope turns out to be 2.5.  Hence, one can conclude that
the temperature dependence of $\mu$ causes the huge discrepancy observed in
the {\it amplitude}.  Note that when the density of states is independent of
the energy, the Fermi energy must remain constant.  Yet, in real simulations,
it changes for each value of $\Gamma$.  
Fortunately, this 5 $ \%$ change in the Fermi temperature does not make 
a noticeable difference in the Fermi fitting.
Therefore, the temperature fitting is robust.
\vskip 0.5 true cm
{\bf 4. Fluctuations of the Center of Mass}

The fluctuations of the center of mass were also measured and plotted as a 
function of $T^3$ for N=200, $\Omega=4$ and $\Omega=8$ and N=100, $\Omega=4$ 
under a sine wave vibration and N=200, $\Omega=4$ and N=100, $\Omega=4$ under 
a triangle wave vibration.(Fig.6)  Once again, this confirms 
the validity of the Fermi statistics, which implies that
$$ <(\Delta z)^2> \propto T^3.$$
The proportionality constant, though, is a way off from the theoretical
value, $D \pi^2/3 \mu^2 \approx 6.4 \bullet
 10^{-3}$.  The actual slope is of the order 1.  
Considering the fact that the amplitude in the center of mass is
off by a factor 10, one can expect the errors in the fluctuations to be
larger, of the order $10^2$.  This is again due to
the sensitivity of the Fermi integral to $\mu$.  A small
change in $\mu$ is magnified in the fluctuations, $<(\Delta z^2)>$, 
resulting in the $10^2$ factor difference.  An additional source of 
discrepancy is due to the fact that the fluctuations of the center of mass
are quite large in the vibrating bed, and
{\it all} the particles down to the ones at the very bottom of
the layer fluctuate in a {\it continuum} space.
This is in contrast to the Fermi particles of a lattice model, 
where most of the particles below the
Fermi surface are locked and are thus,
{\it inactive}.  Hence, while the
average position of the grains appear to obey the Fermi distribution 
function quite well, its actual magnitude in the fluctuations may not.
The surprise here, however, is that the temperature dependence of the 
flutuations still appears to obey the $T^3$ law of the Fermi statistics.

\vskip 1.0 true cm
\noindent {\bf V. Discussion}

There are several issues that should be discussed in connection with the
present work.
\vskip 0.2 true cm
{\it First}, 
the thermodynamic aspect of the theory presented in [1] should be investigated
in more detail in the future.  
Certainly the temperature formula, eq.(1a) and (1b), certainly needs to be
modified and improved. Further, note the obvious fact that
the temperature in the Fermi distribution function 
is not the same as the kinetic temperature, $T_k=m<v^2>/2$.  $T_k$
is zero
when there is no motion, but the temperature T defined in
ref [1] is non zero even though 
there is no motion because of the entropic contribution.  
In this sense, our theory is
similar to that proposed by
Edwards [9].
Edwards and his collaborators proposed a thermodynamic model for an
{\it equilibrium} state by defining a temperature like variable, termed the
compactivity,

$$ X= \frac{\partial V}{\partial S},$$
\noindent 
where V is the total volume of the granular system, and S is its
entropy.  The volume V can be determined easily, but there is
no systematic way of determining the
entropy, S, of a disordered system. Hence,
there is no calculational mechanism to relate the
central parameter of his theory, the compactivity X, to the experimentally
controlled parameters such as $\Gamma$.  Thus, in our opinion
the theory is not closed.  Within his formalism, it is quite a challenge to
systematically compute the entropy of a disordered system.
Recently, the Chicago group
[10] determined
X using a fluctuation formula such as eq.(2b).
Note, however, that this is an {\it experimental} method used to determine
X, and a formula relating X to control parameters is still missing.
\vskip 0.3 true cm
Our theory departs from
Edwards in two ways: First, it is a thermodynamic theory of the {\it 
configurational statistics} of a {\it nonequilibrium dynamical state} or
{\it steady state}.  In this view point, the dynamics of compaction [10,11]
might be viewed as a transition of an unstable state into an equilibrium
state [12].  The temperature equation may be ignored in the hydrodynamic
approach of the convective instability of granular materials [13,14].
Next, the relationship between the temperature variable and
the control parameters was determined by comparing the {\it kinetics}
and {\it thermodynamics}. This provides a closure to Edward's thermodynamic
theory by providing a specific relationship between the temperature or
compactivity and the external control parameters.  
\vskip 0.3 true cm
{\it Second,} the thermodynamic theory
in the presence of gravity with a {\it hard sphere} gas presents some
interesting puzzles and needs to be pursued in the future, notably the
question of whether or not
the temperature is an extensive quantity along
the direction of anisotropy.
For example, consider point particles confined in
a three dimensional
box of size $(L_x,L_y,L_z)$ under gravity along the z direction, for which the
energy level is given by $\epsilon = p^2/2m + mgz$.  One can show
easily that the energy per particle $\bar u$ is given by
$$\bar u\equiv E/N= \frac{1}{N}\frac{\partial lnZ}{\partial\beta} =
\frac{5}{2}kT - f(x) \eqno (3)$$
where $Z=z^N/N!$ is the partition function of the N particle system, 
$z=(2m\pi/\beta)^{3/2}L_yL_x(1-exp(-\beta x))/\beta mg$, and 
$f(x)=xexp(-\beta x)/(1-exp(-\beta x))$ with $x=mgL_z$.  In the limit 
$L_z\rightarrow\infty$, the energy per particle obeys the equipartition
theorem and approaches $5kT/2$.  Further it approaches zero at the
zero temperature limit.  Note also that the total energy E is an
extensive quantity, and the thermodynamic relation between the temperature
and the entropy also is satisfied, namely:
$S/k = ln Z +\beta E$ and 
$$\frac{1}{k}\frac{\partial S}{\partial E} =
\frac{1}{k}(\frac{\partial S}{\partial\beta})(\frac{\partial E}{\partial\beta}
)^{-1} = \beta/k=1/T. \eqno (4)$$

However, for granular materials with finite diameter D(hard spheres), 
the situation becomes
a little different.  In the zero temperature limit, while the
kinetic energy approaches zero, the potential energy is still a function of
disorder and it is {\it not} zero.  For point particles, both the kinetic and
potential energy approach zero because particles can be
compressed indefinitely.  Further, for a hard sphere gas,
the potential energy is not extensive and
thus the temperature T is not intensive.  Consider for example the granular
materials confined in a two dimensional box of width L and the height H,
for which the potential energy $E_o=LH\frac{H}{2} = LH^2/2$.  Now, 
change the box size by a factor two: if the width is increased to 2L,
then the potential energy doubles, i.e:
$$ E=2LH\bullet\frac{H}{2} = 2E_o$$
Hence, the energy is extensive.  On the other hand, if the height is increased 
by a factor 2, then the potential energy increases by a factor 4, i.e:
$$ E= L\bullet 2H\bullet\frac{2H}{2} = 4E_o. \eqno $$
So, the energy is not extensive.  Hence, one finds here that while the
energy does not depend on the way the system is doubled for
point particles, it does for hard spheres with finite diameter.  
The thermodynamic definition of
the temperature with hard spheres in the presence of
gravity does not seem so simple
as in the case of point particles.   This
is not inconsistent with recent experimental result [15], where it
was reported that even though the velocity profiles obey perfect Gaussian,
one needs two different temperatures to describe the
velocity profile of excited grains in a vibrating bed
\vskip 0.3 true cm

{\it Third}, 
comments can be made about the MD code which was used.  It is important 
to note that the MD code used in this work
allows the grains to deform upon contact and the amount of deformation,
in particular along the normal direction, depends on the mass and the
spring constant.  Within this MD code, unlike the
assumption made in the theory that particles are compact and nondeformable,
the temperature does depend weakly on
the {\it mass} of the grain.  More precisely,
when a spring with the spring constant K is displaced
a distance $\Delta z$ by a mass m, the velocity of the mass is given by
$\frac{1}{2}mv^2=\frac{1}{2}K(\Delta z)^2$, i.e. $v=(K/m)^{1/2}$.  Now, note
that in the MD code [8] used in this paper, the normal contact force
at the bottom plate, $F$, has three components. 
$$F = K_n(\Delta r)^m - mg - \gamma_n m_{eff}(\hat v\bullet\hat n)$$
where the first term is the Hertzian contact force [16], the second term is
the gravity term, and the third term is due to the dynamic friction.  Since 
the acceleration of the particle, $a$, is given by 
$a=dv/dt=F_n/m$, and the first term is 
independent of the mass, the bouncing velocity of the particle at the bottom
plate {\it is} a function of the mass for a given $K_n$.
Therefore, the jumping height of the ball {\it does} depend on the mass of the
particle.  While the MD codes that allow such deformation
are based on certain models [17] and 
experiments [18], it is 
not certain at this point whether such mass dependent
dynamics of deformable grains is a realistic modeling of
real granular materials.  Presumably, this ambiguity might have 
had some effect on the discrepancies 
in the amplitude for the center of mass and its fluctuations.
It is highly desirable to carry out the same study
undertaken here with a hard sphere MD code [19] which does not allow
such deformations to occur.
Finally, we also point out that
the microscopic basis of the Fermi statistics has been
recently examined by one of the authors in ref. [20], where it was 
demonstrated that
the crossover from Boltzmann to Fermi statistics as the
vibrational strength decreases arises via the condensation of granular
particles under gravity.  This is another
consequence of the excluded volume interaction of
finite grains that one cannot compress such a system indefinitely,
which is the key to uncovering an interesting
phenomena of 'condensation of grains under gravity.
Putting aside the mathematical aspect of this condensation,
which is {\it mathematically} similar
to Bose condensation, the
underlying physics of this phenomenon is not difficult to
understand.  
If we start with particles at a high temperature and decrease
the temperature slowly, then a portion of particles begins
to settle down from the
bottom, and two phases develop: a fludized region near the top 
and a solid or glass
region near the bottom.  
This is what we term the condensation phenomenon of granular particles,
and in the presence of such two phases, the continuum theory
may pose some problems.  For detailed summary of how the kinetic theory
describes such a condensation phenomenon, see ref.[20].
\vskip 1.0 true cm
\noindent {\bf Acknowledgement}
\vskip 0.2 true cm
We wish to thank many people for their
assistance during the course of this project,
especially  J. Lee for providing us with the Molecular Dynamics code,
and S. Luding for many valuable suggestions on the subtle aspect of MD 
simulations.  We also wish to thank H. Hayakawa for 
constructive criticisms and comments
over the course of this work, as well
as M.Y Choi 
for comments on the thermodynamics of point particles with gravity and
the hard core nature of grains.

\newpage
\noindent {\bf References}
\vskip 0.2 true cm
\noindent [1] H. Hayakawa and D. C. Hong, Phys. Rev. Lett. {\bf 78}, 2764 
(1997)
\vskip 0.2 true cm
\noindent [2] E. Clement and J. Rajchenbach, Europhys. Lett. {\bf 16},
1333 (1991)
\vskip 0.2 true cm
\noindent [3] For the parking lot analogy in slow relaxation of polymer
absorption problem, see
E. Ben-Naim, J. Chem. Phys. {\bf 100}, 6778 (1994) and ref.[14].
\vskip 0.2 true cm
\noindent [4] See., eg., T.L. Hill. {\it Statistical Mechanics} (Dover,
New York, 1987), Chapt.8.
\vskip 0.2 true cm
\noindent [5] S. Warr and J. P. Hansen, Euro. Phys. Lett. Vol. 36, no.8 
(1996)
\vskip 0.2 true cm
\noindent [6] J.P. Hansen and I. R. McDonalds, {\it Theory of Simple
Liquids,} Academic Press, London (1986); See also,  G. Ristow, 
Phys. Rev. Lett. {\bf 79}, 833 (1997).
\vskip 0.2 true cm
\noindent [7] J.M. Luck and A. Mehta, Phys. Rev. E {\bf 48}, 3988 (1993) 
\vskip 0.2 true cm
\noindent [8] See for example: J.A.C. Galls, H. Herrmann, and S. Sokolowski,
Physica(Amsterdam) {\bf 189A}, 437 (1992),
G.H.Ristow, G. Strassburger, and I. Rehberg Phys. Rev. Lett. {\bf 79}, 
833 (1997);  For hard sphere MD codes, see: S. Luding, 'Models and 
simulations of granular materials,' Ph.D. thesis, Albert-Ludwigs University, 
Germany (1994);See also; S. Luding, H. J. Herrmann, and A. Blumen, Phys. Rev.
E {\bf 50}, 3100 (1994); T. Schwagar and T. Poeschel, cond-matt/9711313;
S. McNamara and W. Young, Phys. Fluids. A{\bf 4} (3), 496 (1992)
\vskip 0.2 true cm
\noindent [9] S.F. Edwards and R.B.S. Oakeshott, Physica A {\bf 157}, 1080
(1989); A. Mehta and S.F. Edwards, Physica A(Amsterdam) {\bf 168}, 714 (1990)
For other thermodynamic theories of grains, see: B. Bernu, F, Deylon, 
and R. Dazighi, Phys. Rev. E {\bf 50}, 4551 (1994); J.J.Brey, F. Moreno, and
J.W. Duffy, Phys. Rev. E {\bf 54}, 445 (1996)
\vskip 0.2 true cm
\noindent [10] J.B.Knight, C.G. Fandrich, C.N. Lai, H.M. Jaeger, and
S. R. Nagel, Phys. Rev. E {\bf 51}, 3957 (1995);
E. Nowak, J. Knight, E. Ben-Naim, H.M. Jaeger, and S.R. Nagel, {\it
Density Fluctuations in Vibrated Granular Materials,} Preprint;
\vskip 0.2 true cm
\noindent [11] G. Barker and A. Mehta, Phys. Rev. E {\bf 47}, 184 (1993);
D.C. Hong et al, Phys. Rev. E {\bf 50}, 4123 (1994);
E. Cagloti, V. Loreto, H. J. Herrmann, and M. Nicodemi,
Phys. Rev. Lett. {\bf 79}, 1575 (1997)
\vskip 0.2 true cm
\noindent [12] H. Hayakawa and D. C. Hong, Int. J. Bifurcations and
Chaos, Vol.7, 1159 (1997)
\vskip 0.2 true cm
\noindent [13] M. Bourzutschky and J. Miller, Phys. Rev. Lett. {\bf 74},
2216(1995); H. Hayakawa, S. Yue and D. C. Hong, Phys. Rev. Lett. {\bf 75}, 
2328 (1995); P. Haff, J. Fluid. Mech. {\bf 134}, 401 (1986);
J.T.Jenkins and S.B. Savage, J. Fluid Mech. {\bf 130}; C. Saluena and
T. Poeschel, cond-mat/9807071;
187 (1983); K. Aoki et al, Phys. Rev. E., {\bf 54}, 874 (1998);
Y-h. Taguchi, Phys. Rev. Lett. {\bf 69}, 1367 (1992); 
J. Gallas, H. Herrmann, and S. Sokolowski, {\it ibid}, {\bf 69}, 1371 (1992);
D.C. Hong and S.Yue, Phys. Rev. E {\bf 58}, 4763 (1998)
\vskip 0.2 true cm
\noindent [14] P. Evesque and J. Rajchenbach, Phys. Rev. Lett. {\bf 62},
44 (1989); E. Clement, J. Duran, and J. Rajchenbach, Phys. Rev. Lett.
{\bf 69}, 1189 (1992); Y-h. Taguchi, Europhys. Lett. {\bf 24}, 203(1993);
K. Ichiki and H. Hayakawa, Phys. Rev. E {\bf 52}, 658 (1995); H.K.Pak and R. 
Behringer, Nature(London) {\bf 371}, 231(1994); S. Daoudy, S. Fauve, and 
C. Laroche, Europhys. Lett. {\bf 8}, 621 (1989).
\vskip 0.2 true cm
\noindent [15] J.S. Olafsen and J.S. Urbach, Phys. Rev. Lett. 
{\bf 81}, 4369 (1998);
J. Delour, A. Kudrolli, and J. Gollub, cond-matt/980-6366
\vskip 0.2 true cm
\noindent [16] See for example, {\it Contact Mechanics}, K.L. Johnson,
Cambridge University Press (1985)
\vskip 0.2 true cm
\noindent [17] H. Brandt, J. Appl. Mech. Vol.22, 479 (1955); 
Deresiewicz, {\it ibid}, Vol.25, 402 (1958);
P, J. Digby, {\it ibid} Vol.48, 803 (1981)
\vskip 0.2 true cm
\noindent [18] S. Forester, M. Louge, H. Chang, and K. Allia, Phys.
Fluids. {\bf 6} (3), 1108 (1994) 
\vskip 0.2 true cm
\noindent [19] Jysoo Lee, cond-mat/9606013; 
S. Luding, E. Clement, A. Blumen, J. Rajchenbach, and J. Durna, Phys. Rev.
E. {\bf 50} R1762 (1994)
\vskip 0.2 true cm
\noindent [20] D. C. Hong, ``Condensation of hard spheres under
gravity,'' Cond-matt/9806253.(To appear in Physica A)
\newpage

\noindent {\bf Figure Captions}
\vskip 1.0 true cm
\noindent Fig.1. Density profile for $\Gamma=20$(square) and 
$\Gamma=2$(circle) with N=100 and $\Omega=4$.  For $\Gamma=2$, the inequality 
eq.(3) that insures the validity of the Fermi statistics is satisfied.  
The Fermi statistics breaks down for $\Gamma=20$ and the density profile
is approaches that of the dilute gas with gravity, shown with the  dashed line.
\vskip 0.3 true cm
\noindent Fig.2 Representative density profiles for different $\Gamma$ 's
are plotted for N=100, $\Omega=4$.
\vskip 0.3 true cm
\noindent Fig.3. Comparison between the measured temperatures by MD
and the predicted ones by eq.(2a)and (2b) for (a) sine wave excitations
(b) triangular wave excitations. 
\vskip 0.3 true cm
\noindent Fig.4. The log-log plot of the scaling of the measured temperature, 
$T$, against the controlled parameters (a) frequency - $\omega$, 
(b) gravity - g, and (c) the diameter - D.  The solid lines are a guide for 
the eye and their slopes are: 1.167($\omega$), 0.48(g) and 0.528(D) 
respectively.
\vskip 0.3 true cm
\noindent Fig.5. The center of mass, $<z(T)>$, is plotted for N=100 $\Omega=4$ 
and N=200 $\Omega=4$ for both the sine wave and triangle wave vibration 
and N=200 $\Omega=8$ for the sine wave vibration as a function of $T^2$. The 
straight lines are a guide to the eye.
\vskip 0.2 true cm
\noindent Fig.6. Fluctuations in the center of mass, $<(\Delta z(T)^2>$,is 
plotted for N=100 $\Omega=4$ and N=200 $\Omega=4$ for both the sine wave and 
triangle wave vibration and N=200 $\Omega=8$ for the sine wave vibration as a 
function of $T^3$. The straight line is a guide to the eye.

\newpage
\thispagestyle{empty}
\centerline{\hbox{
\psfig{figure=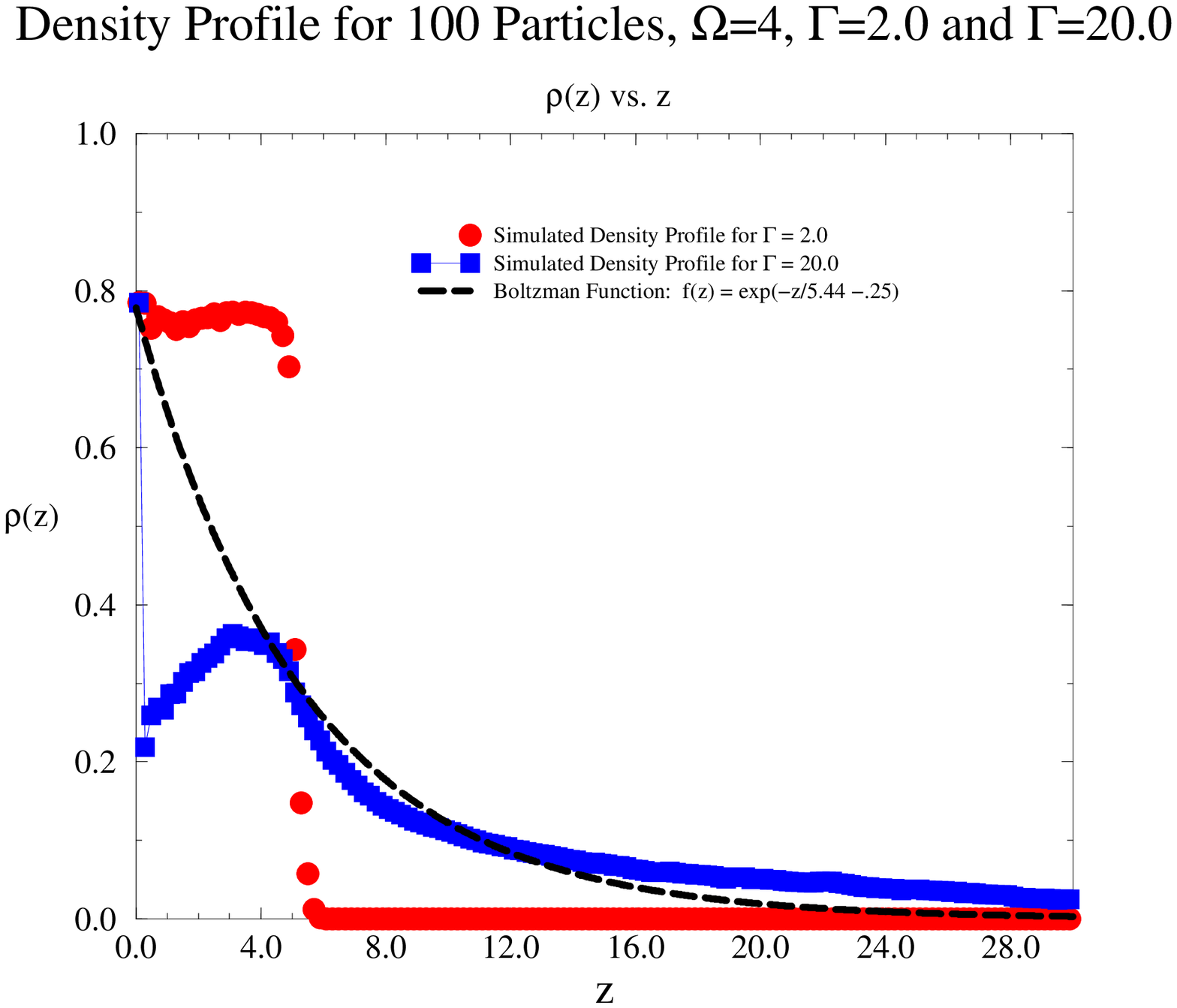}
}}

\newpage
\thispagestyle{empty}
\centerline{\hbox{
\psfig{figure=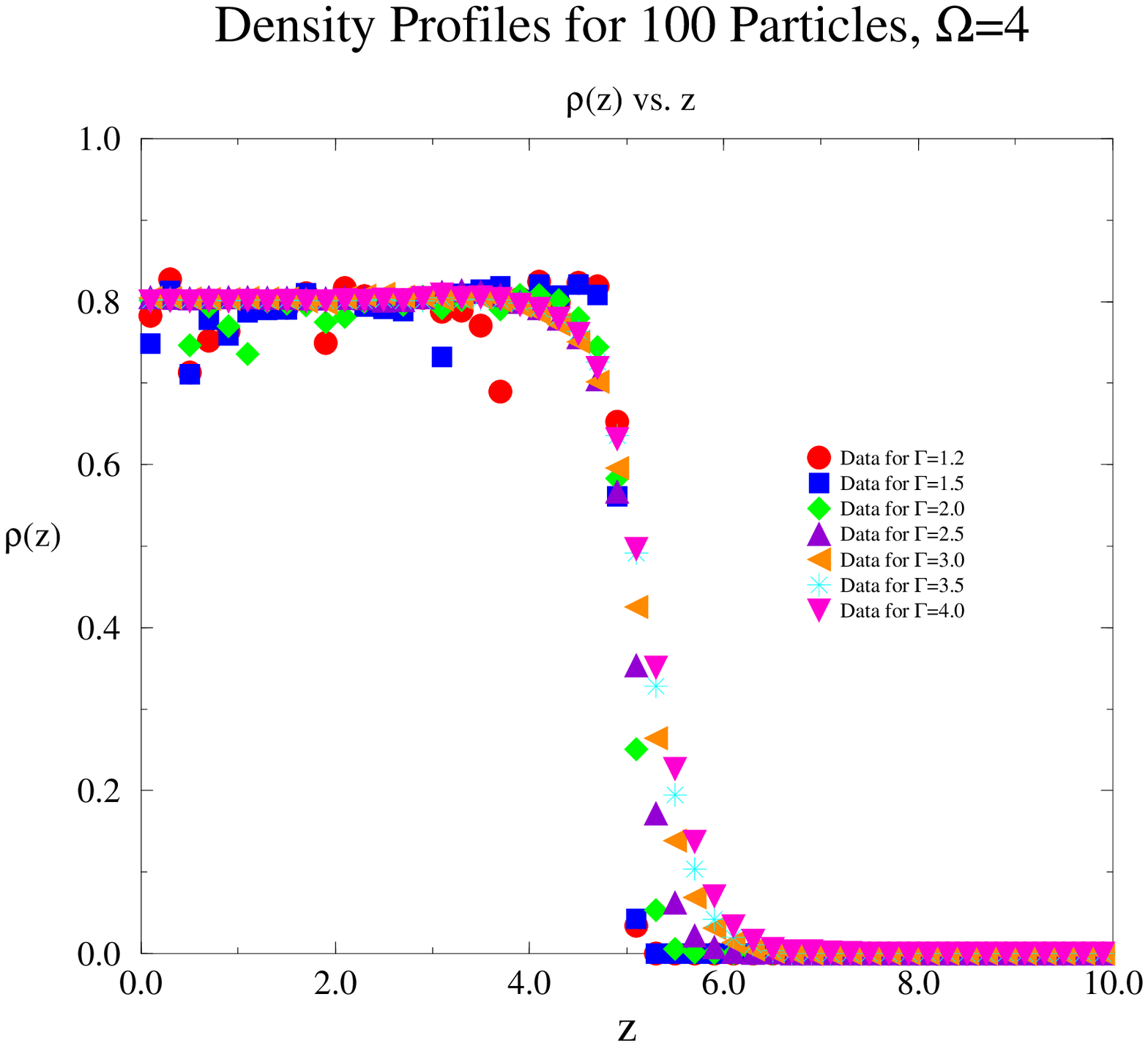}
}}

\newpage
\thispagestyle{empty}
\centerline{\hbox{
\psfig{figure=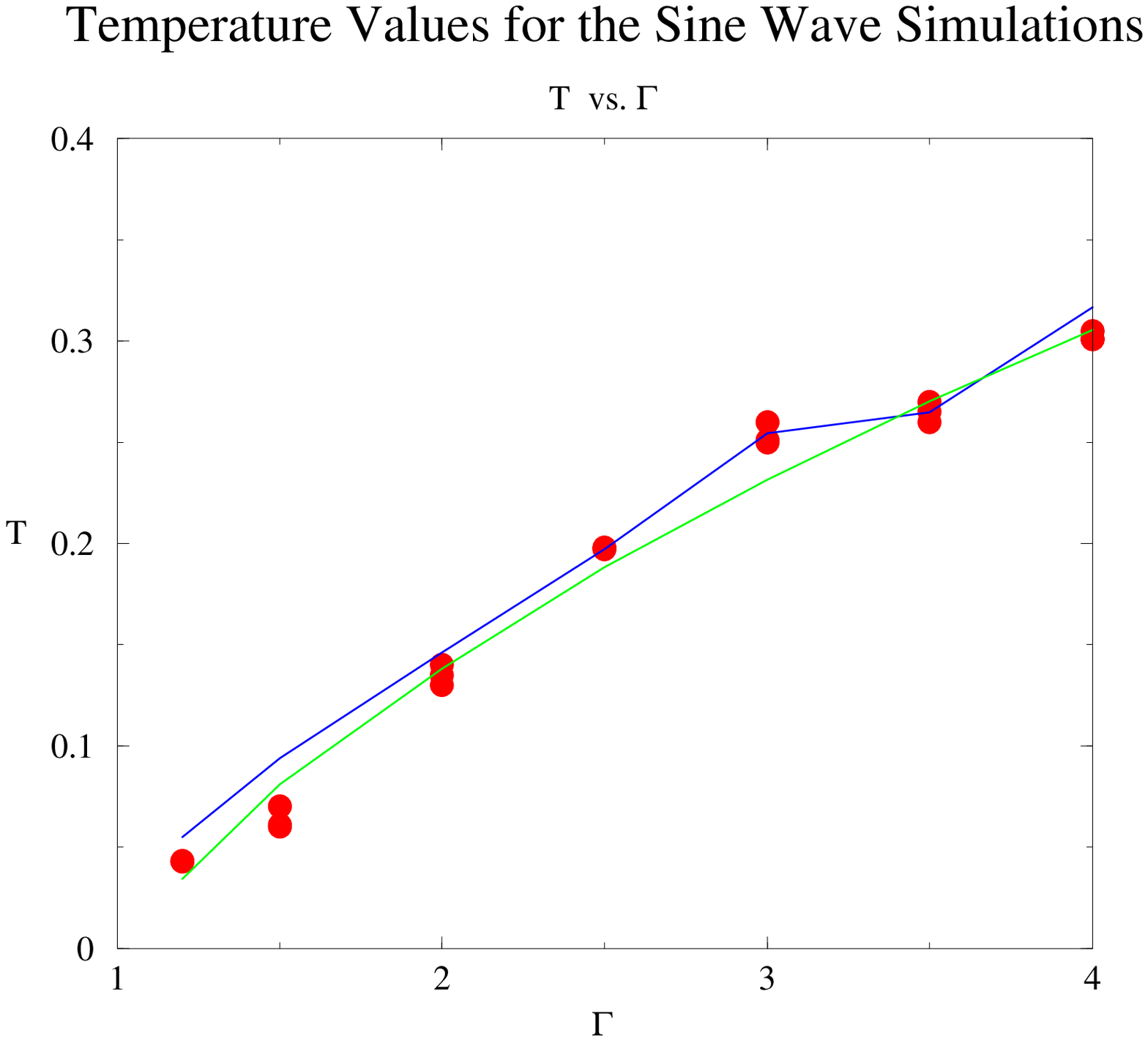}
}}

\newpage
\thispagestyle{empty}
\centerline{\hbox{
\psfig{figure=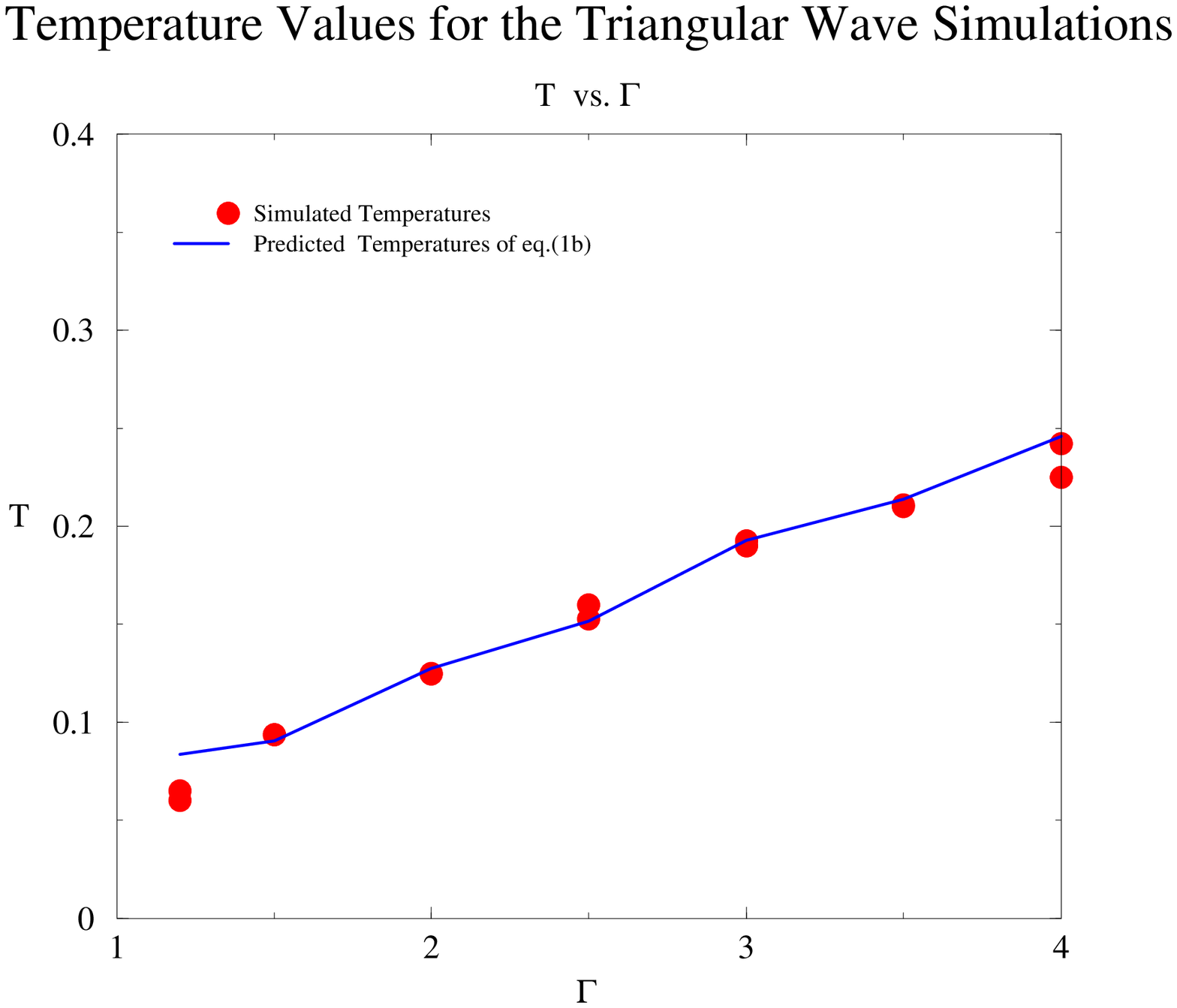}
}}

\newpage
\thispagestyle{empty}
\centerline{\hbox{
\psfig{figure=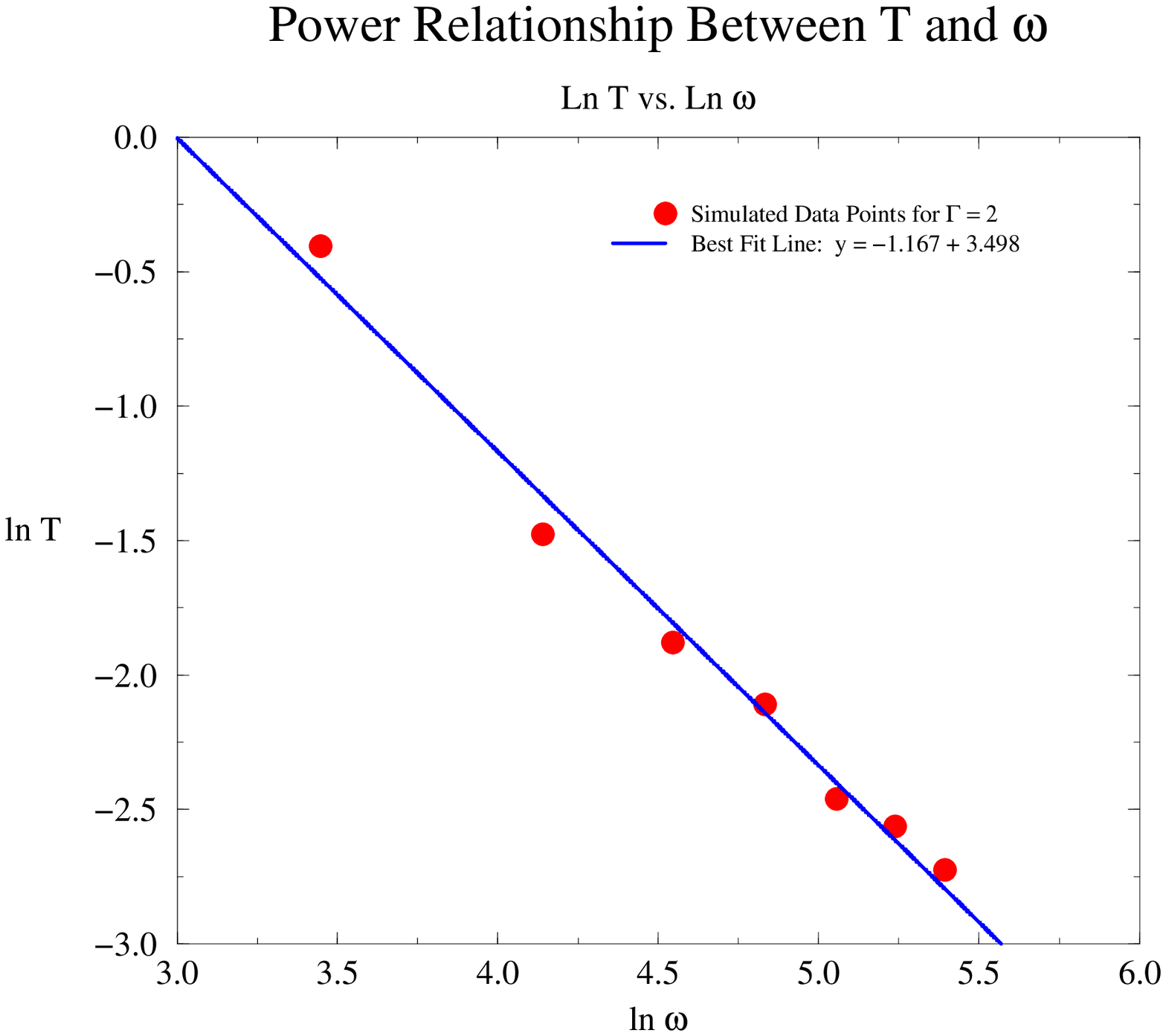}
}}

\newpage
\thispagestyle{empty}
\centerline{\hbox{
\psfig{figure=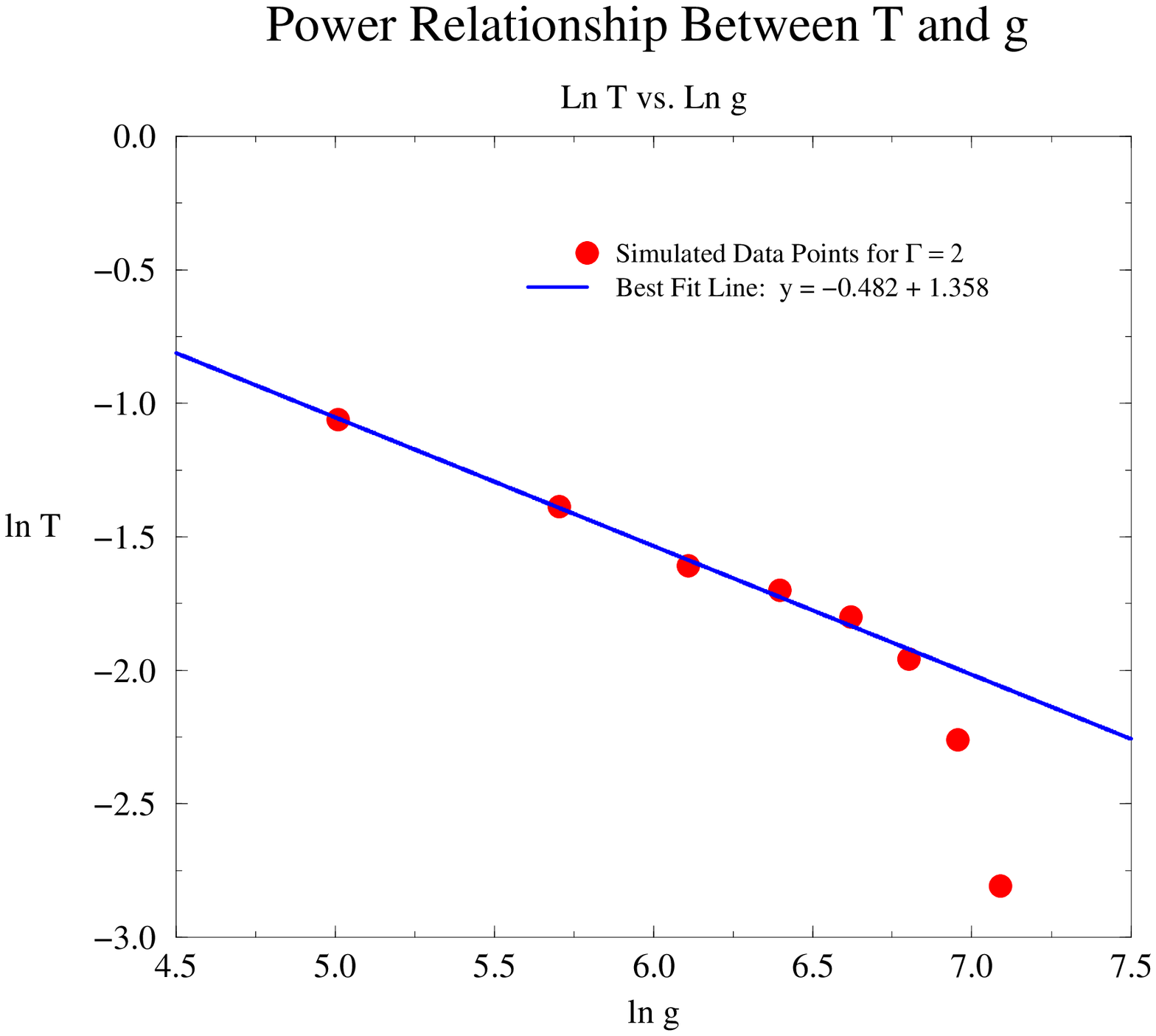}
}}

\newpage
\thispagestyle{empty}
\centerline{\hbox{
\psfig{figure=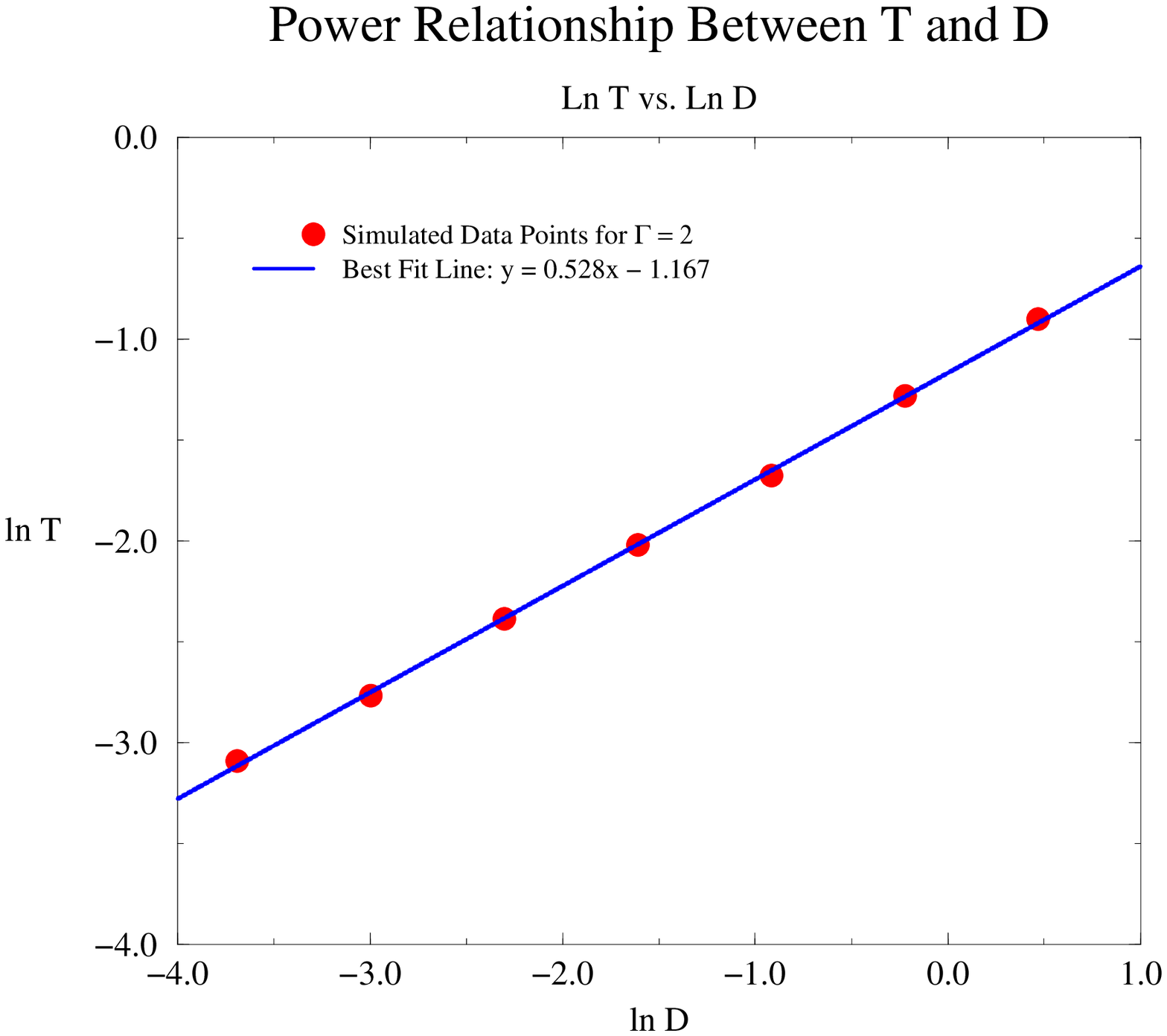}
}}

\newpage
\thispagestyle{empty}
\centerline{\hbox{
\psfig{figure=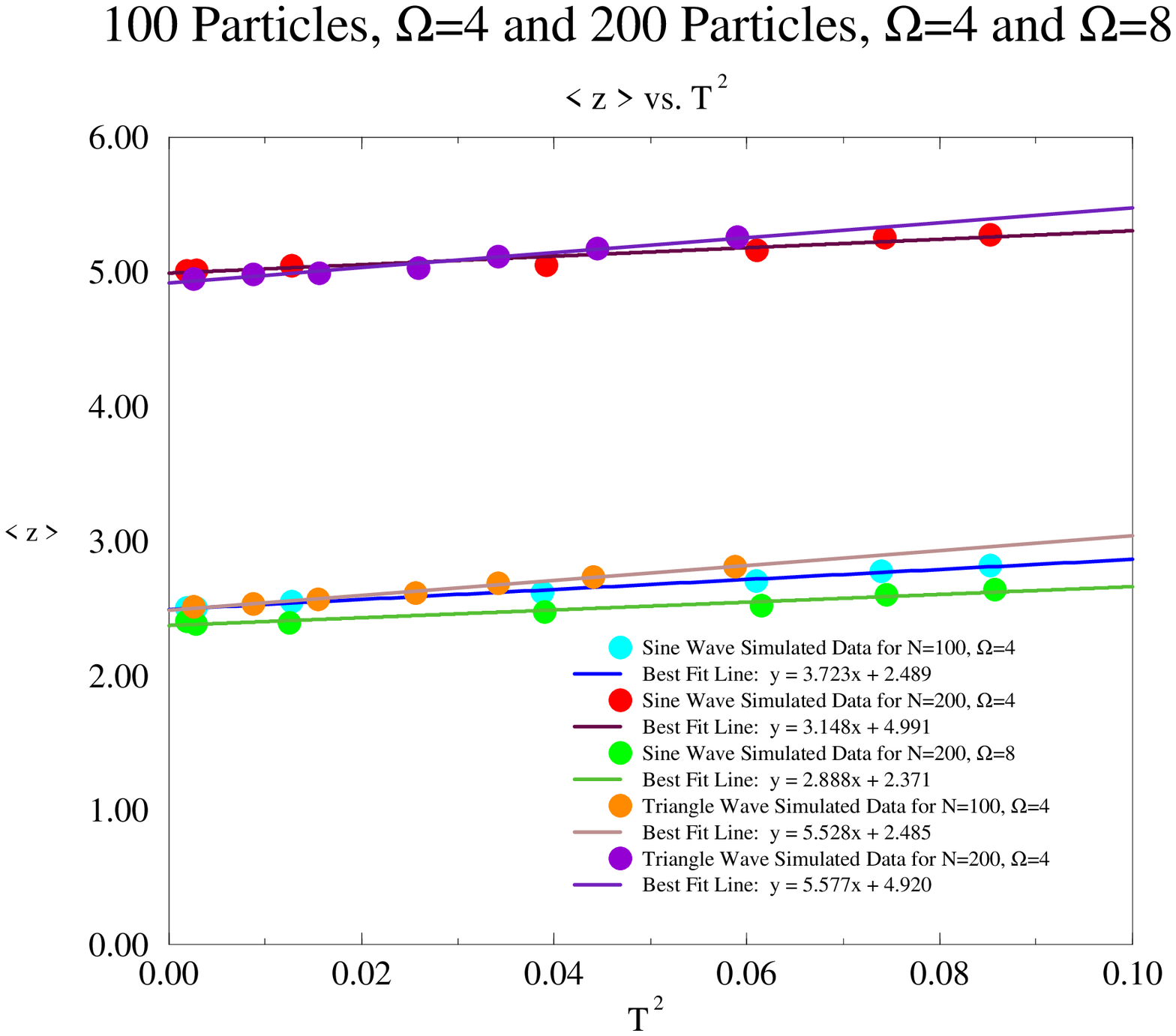}
}}

\newpage
\thispagestyle{empty}
\centerline{\hbox{
\psfig{figure=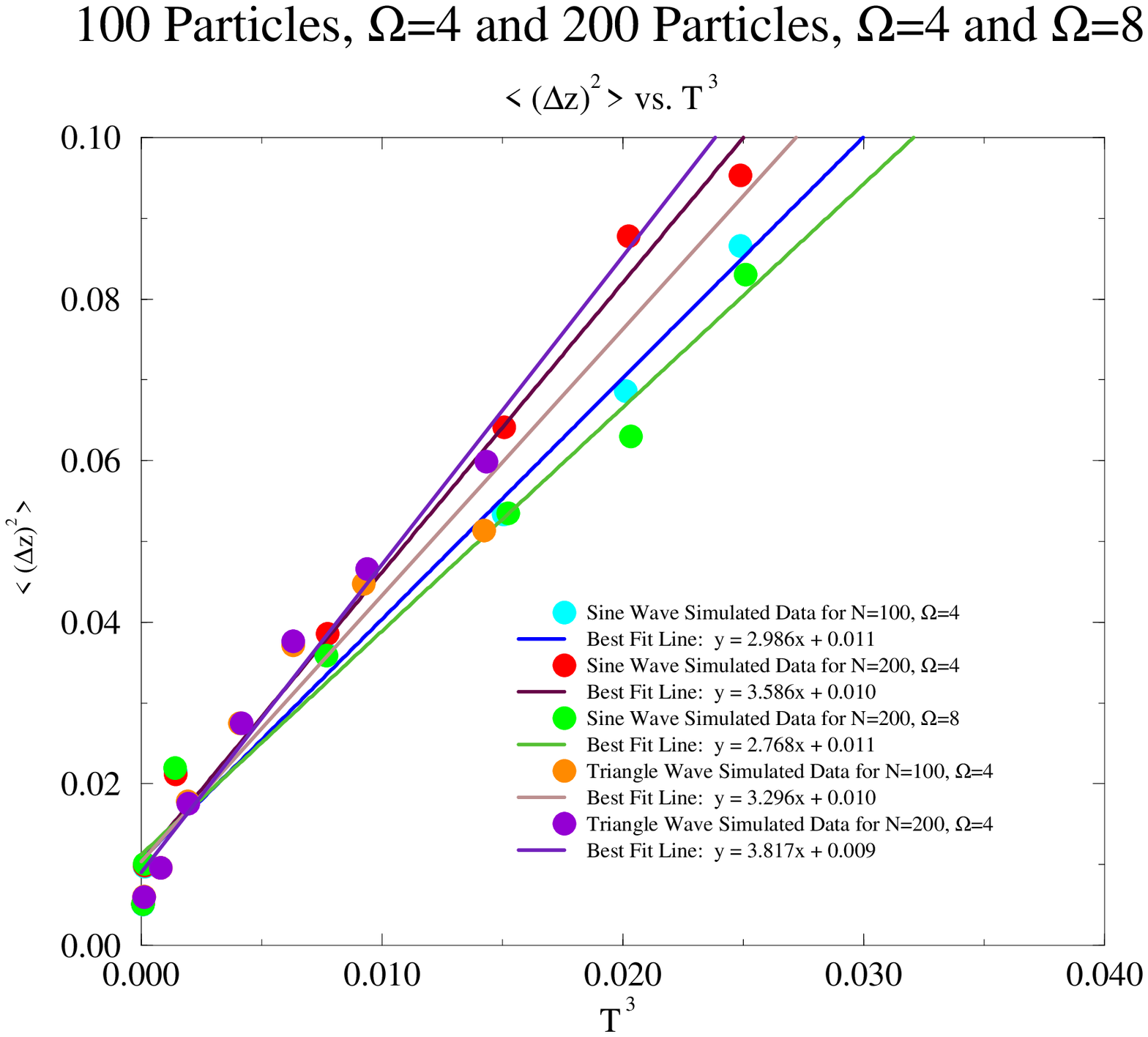}
}}

\end{document}